\documentclass{elsart}
\usepackage{amssymb,epsfig,graphicx,amsmath}

\journal{Phys. Lett. A}
\begin{document}
\begin{frontmatter}

\title{Resonances in the one-dimensional
Dirac equation in the presence of a point interaction and a constant
electric field}
\author[IVIC]{Luis Gonz\'alez-D\'{\i}az},
\ead{lugonzal@pion.ivic.ve}
\author[IVIC]{V\'{\i}ctor M. Villalba\corauthref{cor}}
\corauth[cor]{Corresponding author.} \ead{villalba@ivic.ve}

\address[IVIC]{Centro de F\'{\i}sica
Instituto Venezolano de Investigaciones Cient\'{\i}ficas, IVIC \\
Apdo 21827, Caracas 1020-A, Venezuela}

\begin{abstract}
We show that the energy spectrum of the one-dimensional Dirac
equation in the presence of a spatial confining point interaction
exhibits a resonant behavior when one includes a weak electric
field. After solving the Dirac equation in terms of parabolic
cylinder functions and showing explicitly how the resonant behavior
depends on the sign and strength of the electric field, we derive an
approximate expression for the value of the resonance energy in
terms of the electric field and delta interaction strength.
\end{abstract}

\end{frontmatter}

\section{Introduction}

Supercritical effects are perhaps one of the most interesting
phenomena associated with the charged vacuum in the presence of
strong electric fields \cite{Greiner,Rafelski} The study of
supercritical effects induced by strong vector potentials goes back
to the pioneering works of Pieper and Greiner \cite{Pieper},
Zeldovich and Popov \cite{Zeldovich}
 among others. The idea behind
supercriticality is to have positron emission induced by the
presence of very strong attractive vector potentials. The phenomenon
can be described as follows: the energy level of an unoccupied bound
state dives into the negative energy continuum. i.e., an electron of
the Dirac sea is trapped by the potential, leaving  a positron that
escapes to infinity. \ The
electric field responsible for supercriticality should be stronger than $%
2m_{e}c^{2}$, which is the value of the gap between the negative and
positive energy continua. \ Such strong electric fields could be
produced in heavy-ion collisions \cite{Greiner,Greiner2}. A rigorous
mathematical study of the behavior of the Dirac energy levels near
the continuum spectrum and the problem of spontaneous pair creation
has been carried out by \u{S}{e}ba \cite{PS:86}, Klaus \cite{Klaus}
and Nenciu \cite{Nenciu} among others.

In order to get a deeper understanding of the mechanism responsible
for supercriticality and for the resonant peaks appearing in the
energy spectrum when supercritical fields are present, we proceed to
work with a vector point interaction in the presence of a constant
electric field. Point interactions potentials may be used to
approximate, in a simple way, more complex short-ranged potentials.
Among the advantages of working with confining delta interactions we
should mention that, they only possess a single bound state and the
treatment of the interacting potential reduces to a boundary
condition. The study of bound states of the relativistic wave
equation in the presence of point interactions is a problem that has
been carefully discussed in the literature
\cite{Sutherland,McKellar,McKellar2,FD:89,Albeverio}. The
one-dimensional Dirac equation in the presence of a vector point
delta interaction has also been a subject of study in the search of
supercritical effects induced by attractive potentials
\cite{Greiner,Loewe}. Soon after the publication of the paper by
Loewe and Sanhueza \cite{Loewe}, Nogami et al \cite {Nogami} pointed
out that supercritical effects are also absent in a class of
non-local separable potentials in one dimension.

Since we are interested in studying the mechanism of positron
production by supercritical fields, we proceed to analyze the
resonant behavior of the energy when a bound state dives into the
negative continuum. This resonant behavior is associated with the
appearance of simple poles of the resolvent  on the second sheet at
a position very near the real axis \cite{Reed}.

The method of complex eigenvalues (Gamow vectors) was introduced in
quantum mechanics by Gamow \cite{GG:28} in connection with the
theory of Alpha decay. Titchmarsch \cite{ET:58} and Barut
\cite{WB:62} demonstrated an application,  in the framework of
non-relativistic quantum mechanics, for the Gamow vector method to
the problem of an attractive delta interaction $\delta (r)$ of strength $%
-\cot \alpha $ with a weak electric  field term associated with the
potential $V(r)=-\lambda r.$   They found that the Schr\"{o}dinger
equation with a weak electric field exhibits a continuous spectrum
from $-\infty $ to $+\infty ,$ and  a resonance at $E^{'}$ in the
vicinity
of $E_{{0}}-\frac{1}{2}\lambda \tan \alpha ,$ where $%
E_{0}=-\cot ^{2}\alpha $ is the energy of the unperturbed state. The
eigenvalue $E^{'}$
is a complex number in the lower half plane. \ According to Barut, the Schr%
\"{o}dinger equation with the potential $-\lambda r$ describes a
system that tries to form a bound state that ``dissolves itself" in
the presence of the continuous spectrum. In this Letter, using \ the
idea developed by Titchmarsch \cite{ET:58}, we find the energy
spectrum of the one-dimensional Dirac equation with boundary
conditions associated with a vector Dirac delta interaction and a
constant electric field whose strength is weak, and therefore it
produces a perturbative effect on the delta energy spectrum. We find
that in this case the energy spectrum exhibits a resonance due to
supercriticality.

The Letter is structured as follows: In Section 2, we solve the
one-dimensional Dirac equation in the presence of an attractive
$\delta$ potential and a constant electric field. In Section 3, we
compute the energy resonances and show how they depend on the
electric field strength. We also derive an approximate analytic
expression for the energy resonances. Finally, in Section 4 we
summarize our conclusions.

\section{The one-dimensional Dirac equation}

In this section we will consider the 1+1 Dirac equation in the presence of
the attractive vector point interaction potential represented by $%
eV(x)=-g_{v}\delta (x),$ and a constant electric field associated
with the potential $eV(x)=\lambda x$. \ The Dirac equation,
expressed in natural units\ ($\hslash =c=1$) can be written in the
form \cite{Greiner1}
\begin{equation}
\left( i\gamma ^{\mu }(\frac{\partial }{\partial x^{\mu }}-ieA_{\mu
})-m\right) \Psi =0,  \label{1}
\end{equation}
where $A_{\mu}$ is the vector potential, $e$ is the charge and $m$
is the mass of the electron. The Dirac matrices $\gamma^{\mu}$
satisfy the commutation relation $\left\{ \gamma
^{\mu },\gamma ^{\nu }\right\} =2\eta ^{\mu \nu }$ with $\eta ^{\mu \nu }=%
\mathrm{diag}(1,-1).$ Since we are working in 1+1 dimensions, we choose to
work in a two-dimensional representation of the \ Dirac matrices
\begin{equation}
\gamma ^{0}=\sigma _{3},\gamma ^{1}=-i\sigma _{2}  \label{2}.
\end{equation}
Substituting the representation matrix representation (\ref{2}) into
Eq. (\ref{1}), and taking into account that the potential
interaction does not depend on time, we obtain

\begin{equation}
\{-i\sigma _{{1}}\frac{d}{dx}+\left( \lambda x-E\right) +m\sigma _{{3}}\}%
\mathsf{X}(x)=0,  \label{dirac}
\end{equation}
with $\Psi =\sigma _{3}\mathsf{X.}$, and
\begin{equation}
\mathsf{X}(x)=
\begin{pmatrix}
X_{{1}} \\
X_{{2}}
\end{pmatrix}
,
\end{equation}
with the boundary conditions at $x=0$
\begin{equation}
\begin{split}
X_{{1}}(0^{{+}})& =X_{{1}}(0^{{-}})\cos {g_{{v}}}-iX_{{2}}(0^{{-}})\hspace{%
0.1cm}\mathrm{\sin }{g_{{v}}}, \\
X_{{2}}(0^{{+}})& =-iX_{{1}}(0^{{-}})\hspace{0.1cm}\mathrm{\sin }{g_{{v}}}+X_{%
{2}}(0^{{-}})\cos {g_{{v}}}.
\end{split}
\label{dominio}
\end{equation}
The above conditions (\ref{dominio}) describe a the point vector
potential interaction of strength $g_{{v}}$ \cite{FD:89}.

Equation (\ref{dirac}) is equivalent to the system of equations
\begin{equation}
\left( m+\lambda x-E\right) X_{{1}}-i\frac{dX_{2}}{dx}=0,
\label{pdirac}
\end{equation}
\begin{equation}
i\frac{dX_{1}}{dx}+\left( m-\lambda x+E\right) X_{{2}}=0,
\label{sdirac}
\end{equation}
Introducing the new functions $\Omega _{{1}}$ and $\Omega _{{2}}$
\begin{equation}
X_{{1}}=\Omega _{{1}}+i\Omega _{{2}},\quad X_{{2}}=\Omega _{{1}}-i\Omega _{{2%
}},  \label{rela}
\end{equation}
we obtain that the system of equations (\ref{pdirac}) -(\ref{sdirac})
reduces to the form
\begin{align}
\frac{d\Omega _{1}}{dx}+i\left( \lambda x-E\right) \Omega _{{1}}-m\Omega _{{2%
}}& =0,  \label{podirac} \\
\frac{d\Omega _{{2}}}{dx}-i\left( \lambda x-E\right) \Omega _{{2}}-m\Omega _{%
{1}}& =0,  \label{sodirac}
\end{align}
which is more tractable in the search of exact solutions. Substituting (\ref
{podirac}) into (\ref{sodirac}) we obtain the second-order differential
equation
\begin{equation}
\frac{d^{2}\Omega _{{1}}}{dx^{2}}\,+\{i\lambda +\left( \lambda
x-E\right) ^{2}-m^{2}\}\,\Omega _{{1}}=0.  \label{ecua}
\end{equation}

Looking at the asymptotic behavior of the parabolic cylinder
functions $D_{\nu}(z)$ \cite{HB:53} we obtain that the regular
solutions, for $\lambda>0$, of Eq. (\ref{ecua}) belonging to
$\mathcal{L}^{2}(-\infty ,0)$ and $\mathcal{L}^{2}(0,\infty )$
($Im\,E>0$), respectively are
\begin{equation}
\begin{split}
\Omega _{{1}}^{-}(x)& =AD_{{-\rho -1}}\left( \sqrt{\frac{2}{\lambda }}e^{{-i%
\frac{\pi }{4}}}\left( \lambda \,x-E\right) \right)  \\
\Omega _{1}^{+}(x)& =BD_{\rho }\left( \sqrt{\frac{2}{\lambda }}e^{{i\frac{%
\pi }{4}}}\left( \lambda \,x-E\right) \right) ,
\end{split}
\label{sol1}
\end{equation}
where $D_{{\rho }}$ y $D_{{-\rho -1}}$ are parabolic cylinder
functions \cite{HB:53}, $\rho =\frac{im^{{2}}}{2\lambda }$, \ and
$A$ and $B$ are constants.

Inserting (\ref{sol1}) into (\ref{podirac}) and using the recurrence
relations for the parabolic cylinder functions \cite{HB:53}, we
obtain
\begin{equation}
\begin{split}
\Omega _{2}^{-}(x)& =i\frac{\sqrt{2\lambda }}{m}e^{{i\frac{\pi }{4}}}AD_{{%
-\rho }}\left( \sqrt{\frac{2}{\lambda }}e^{{-i\frac{\pi }{4}}}\left( \lambda
\,x-E\right) \right),  \\
\Omega _{2}^{+}(x)& =i\frac{m}{\sqrt{2\lambda }}e^{{i\frac{\pi }{4}}}BD_{{%
\rho -1}}\left( \sqrt{\frac{2}{\lambda }}e^{{i\frac{\pi }{4}}}\left(
\lambda \,x-E\right) \right).
\end{split}
\label{sol2}
\end{equation}
From (\ref{sol1}), (\ref{sol2}) and (\ref{rela}), we have that the
components of the spinor solution for $x<0$ are

\begin{equation}
\begin{split}
X_{1}^{-}(x)& =A\left[ D_{{-\rho -1}}\left( \sqrt{\frac{2}{\lambda }}e^{{-i%
\frac{\pi }{4}}}\left( \lambda \,x-E\right) \right) -\frac{\sqrt{2\lambda }}{%
m}e^{{i\frac{\pi }{4}}}D_{{-\rho }}\left( \sqrt{\frac{2}{\lambda }}e^{{-i%
\frac{\pi }{4}}}\left( \lambda \,x-E\right) \right) \right],  \\
X_{{2}}^{-}(x)& =A\left[ D_{{-\rho -1}}\left( \sqrt{\frac{2}{\lambda }}e^{{-i%
\frac{\pi }{4}}}\left( \lambda \,x-E\right) \right) +\frac{\sqrt{2\lambda }}{%
m}e^{{i\frac{\pi }{4}}}D_{{-\rho }}\left( \sqrt{\frac{2}{\lambda }}e^{{-i%
\frac{\pi }{4}}}\left( \lambda \,x-E\right) \right) \right],
\end{split}
\label{izq}
\end{equation}
analogously, we have that, for $x>0$, the spinor $\mathsf{X}$ has
the components:

\begin{equation}
\begin{split}
X_{{1}}^{+}(x)& =B\left[ D_{{\rho }}\left( \sqrt{\frac{2}{\lambda }}e^{{i%
\frac{\pi }{4}}}\left( \lambda \,x-E\right) \right) -\frac{m}{\sqrt{2\lambda
}}e^{{i\frac{\pi }{4}}}D_{{\rho -1}}\left( \sqrt{\frac{2}{\lambda }}e^{{i%
\frac{\pi }{4}}}\left( \lambda \,x-E\right) \right) \right],  \\
X_{{2}}^{+}(x)& =B\left[ D_{{\rho }}\left( \sqrt{\frac{2}{\lambda }}e^{{i%
\frac{\pi }{4}}}\left( \lambda \,x-E\right) \right) +\frac{m}{\sqrt{2\lambda
}}e^{{i\frac{\pi }{4}}}D_{{\rho -1}}\left( \sqrt{\frac{2}{\lambda }}e^{{i%
\frac{\pi }{4}}}\left( \lambda \,x-E\right) \right) \right].
\end{split}
\label{der}
\end{equation}

Inserting (\ref{izq}) and (\ref{der}) in (\ref{dominio}), we obtain that the
equation for the energy eigenvalues $E$ has the form:
\begin{equation}
\begin{split}
2\lambda D_{{\rho }}\left( \sqrt{\frac{2}{\lambda }}e^{{i\frac{\pi }{4}}%
}E\right) & D_{{-\rho }}\left( {-\sqrt{\frac{2}{\lambda }}}e^{{-i\frac{\pi }{%
4}}}E\right)  \\
& -m^{{2}}D_{{\rho -1}}\left( \sqrt{\frac{2}{\lambda }}e^{{i\frac{\pi }{4}}%
}E\right) D_{{-\rho -1}}\left( {-\sqrt{\frac{2}{\lambda }}}e^{{-i\frac{\pi }{%
4}}}E\right) e^{{-2ig_{{v}}}}=0.
\end{split}
\label{urav}
\end{equation}

The regular solutions of Eq. (\ref{ecua}), for $\lambda<0$, can be
obtained after interchanging the roles of $\Omega_{1}^{-}(x)$ and
$\Omega_{1}^{+}(x)$  and of $\Omega_{2}^{-}(x)$ and
$\Omega_{2}^{+}(x)$ in Eq. (\ref{sol1}) and Eq. (\ref{sol2})
respectively; therefore Eq. (\ref{urav}) permits us to compute the
energy resonances for positive as well as negative values of the
electric field strength $\lambda$

\section{Energy resonances}

It is not an easy task to solve Eq. (\ref{urav}), nevertheless it is
possible to derive an analytic approximation for the energy
eigenvalues. In order to do this approximation we consider the
effect of the electric field in the vicinity of zero, i.e., where
the delta interaction is located. We insert the spinor solutions
obtained using this approach into (\ref {dominio}), obtaining in
this way an eigenvalue equation, whose roots are a
good approximation to those of (\ref{urav}). The approximate equation for $%
\Omega _{{1}}$ has the form
\begin{equation}
\frac{d^{2}\Omega _{{1}}}{dx^{2}}\,+\{i\lambda
+E^{2}-m^{2}\}\,\Omega _{{1}}=0.  \label{a}
\end{equation}
The solution of Eq. (\ref{a}) that, for $x>0$, and $\lambda>0$
exhibit a regular asymptotic behavior is
\begin{equation}
\Omega _{1}^{+}=A\exp (-i\sqrt{i\lambda +E^{2}-m^{2}}x). \label{b}
\end{equation}
Substituting the solution (\ref{b}) into Eq. (\ref{podirac}) we obtain
\begin{equation}
\Omega _{2}^{+}=-iA\frac{\sqrt{i\lambda +E^{2}-m^{2}}+E}{m}\exp
(-i\sqrt{i\lambda +E^{2}-m^{2}}x),  \label{d}
\end{equation}
and, analogously, we have that for $x<0,$ the corresponding
solutions are
\begin{equation}
\Omega _{1}^{-}=B\exp (i\sqrt{i\lambda +E^{2}-m^{2}}x), \label{c}
\end{equation}
\begin{equation}
\Omega _{2}^{-}=-iB\frac{-\sqrt{i\lambda +E^{2}-m^{2}}+E}{m}\exp
(i\sqrt{i\lambda +E^{2}-m^{2}}x).  \label{e}
\end{equation}
Substituting the right $\Omega _{1,2}^{+}$ and left $\Omega
_{1,2}^{-}$\ spinor components into Eq. (\ref{dominio}), we find
that, for $g_{v}=-\pi $, the energy state sinks into the negative
energy continuum, exhibiting a resonant behavior that depends on the
electric field strength as:

\begin{equation}
E\left( \lambda \right) \approx -\left( m+\frac{\lambda ^{{2}}}{8m^{{3}}}%
\right) +i\frac{\lambda }{2m},  \label{spectrum}
\end{equation}
where we have assumed that $\lambda$ is positive and small compared
to $m$. It should be noticed that, when we turn off the electric
perturbation, we recover the
energy eigenvalue $E(0)=-m$  for a point vector interaction with strength $%
g_{{v}}=-\pi $ \cite{FD:89}.  It is worth mentioning the vanishing
of the first order perturbative correction to the supercritical
energy $E=-m$ as a result of the presence of the electric field
interaction $\lambda$. This behavior can also be observed looking at
the real part of the resonant energy given by Eq. (\ref{spectrum}).
An approximate expression for the energy resonances for negative
values of $\lambda$ can be obtained after substituting $\lambda$ by
$-\lambda$ in Eq. (\ref{spectrum}).

We can try to obtain an approximate solution to the full Eq.
(\ref{ecua}) after making a series expansion of the function $\
i\lambda +\left( \lambda x-E\right) ^{2}-m^{2}$ around zero. In this
way, we have that the approximate function  $\Omega _{{1}}$
satisfies the second order differential equation
\begin{equation}
\frac{d^{2}\Omega _{{1}}}{dx^{2}}\,+\{i\lambda +E^{2}-2\lambda
Ex-m^{2}\}\,\Omega _{{1}}=0.  \label{f}
\end{equation}
The regular solutions $\Omega _{{1}}^{-}$ and for $x<0,$ and  $\Omega _{{1}%
}^{+}$ for $x>0$ can be expressed, for $\lambda>0$, in terms of Hankel functions $%
H_{1/3}^{1}(z)$ and $H_{1/3}^{2}(z)$ \cite{Abramowitz} as follows
\begin{equation}
\Omega _{{1}}^{-}=Az^{1/3}H_{1/3}^{1}(z),  \label{g}
\end{equation}
\begin{equation}
\Omega _{{1}}^{+}=Bz^{1/3}H_{1/3}^{2}(z),  \label{h}
\end{equation}
where $A$ and $B$ are constants and $z$ is given by the expression
\begin{equation}
z=-\frac{\left( i\lambda +E^{2}-m^{2}-2E\lambda x\right) ^{3/2}}{3\lambda E}
\label{i}
\end{equation}
Substituting the solutions (\ref{g}) and (\ref{h}) for $\Omega
_{{1}}^{-}$ and $\Omega _{{1}}^{+}$ into Eq. (\ref{podirac}) we
obtain the corresponding expressions for $\Omega _{{2}}^{-}$ and
$\Omega _{{2}}^{+}.$ \ Taking into account the relation between
$X_{1}$, $X_{2}$ and $\Omega _{1},$ $\Omega _{2} $ (\ref{rela}), and
inserting the lower and upper components of $\mathsf{X}$ into the
boundary condition (\ref{dominio}), we find that an approximate
equation for the energy spectrum is:

\begin{equation}
\left( X_{{2}}(0^{{-}})X_{{1}}(0^{{+}})-X_{{2}}(0^{{+}})X_{{1}}(0^{{-}%
})\right) \cos g_{v}+i\left( X_{{2}}(0^{{+}})X_{{2}}(0^{{+}})-X_{{1}}(0^{{+}%
})X_{{1}}(0^{{-}})\right) \sin g_{v}=0  \label{energ}
\end{equation}
It is worth noting that the energy spectrum obtained from Eq. (\ref{energ}%
) with $\Omega _{{1}}^{-}$ and $\Omega _{{1}}^{+}$ given by Eqs.
(\ref{g}) and (\ref{h}) gives a result comparable to the one
obtained using the relation Eq. (\ref{spectrum})
\begin{figure}[tbp]
\begin{center}
\includegraphics[width=10cm]{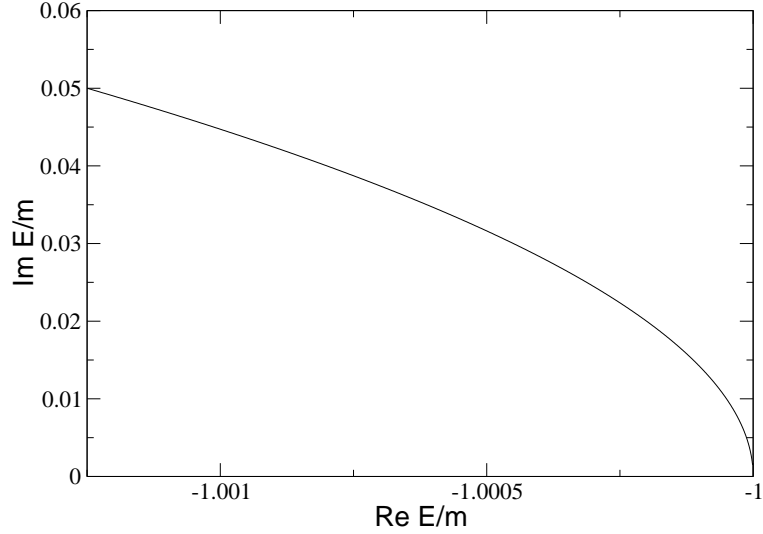}
\end{center}
\caption{Dependence of the real and imaginary parts of the energy on
the electric field strength for $0<\lambda<0.1m^2$.}
\end{figure}
Fig. 1 shows how the real and imaginary part of the energy depend on
the  parameter $\lambda $ when a point vector interaction of
strength $g_{{v}}=-\pi $ is applied. The energy exhibits a resonant
behavior produced by the electric field.

The inverse of the imaginary part of Eq. (\ref{spectrum}) permits
one to determine the time \cite{Newton,Goldberger} that the state
stays in negative continuum.  The stronger the electric field
perturbation, the shorter the mean life of the resonance
\cite{Newton}. An interesting result that can be observed from Eq.
(\ref{spectrum}) and it is also depicted in Fig. 1 is that the
imaginary part of the resonant energy behaves as a square root with
respect to the real part of the energy.

\section{Concluding remarks}

In this article we have shown that the presence of a weak electric
field plays a crucial role in the appearance of a resonant energy
state in the one-dimensional Dirac equation in the field of an
attractive vector delta interaction. This phenomenon is analogous to
the Stark effect \cite{Galindo}, where a perturbative constant
electric field produces a shifting and broadening of the energy
levels of the hydrogen atom. It also parallels the autoionization
process, where the presence electron-electron interaction term
induces a self-ionization process and levels decay like in the Auger
effect \cite{Reed,Galindo}. Such analogies seem to indicate that,
for resonant effects, the shape of the perturbative potential is
sometimes more important than the strength of the potential field.

We have also shown that, the one-dimensional Dirac equation in the
field of an attractive vector delta interaction exhibits a resonant
supercritical  behavior when one introduces a  potential of the form
$\lambda x$. Since this effect is not present when one consider the
sole vector interaction $-g_{v}\delta (x)$ \cite{FD:89}, we conclude
that it is necessary the presence of the electric field  in order to
sink the delta bound state into the negative continuum.

\section{Acknowledgments}
We thank Dr. Ernesto Medina for reading and improving the
manuscript. We also wish to express our gratitude to the anonymous
referees for their critical remarks. This work was supported by
FONACIT under project G-2001000712.

\end{document}